\documentclass[amsmath,superscriptaddress,showpacs,aps,prb,twocolumn]{revtex4-1}
\usepackage[colorlinks,linkcolor=blue,anchorcolor=blue,citecolor=blue,urlcolor=blue]{hyperref}
\usepackage{amsmath}
\usepackage{amssymb}
\usepackage{graphicx}
\usepackage{color}
\usepackage{bm}

\begin{document}

\title{Topological state transfers in cavity-magnon system}

\author{ Xi-Xi Bao}

\affiliation{Lanzhou Center for Theoretical Physics, Key Laboratory of Theoretical Physics of Gansu Province, Lanzhou University, Lanzhou, Gansu 730000, China}

\author{Gang-Feng Guo}

\affiliation{Lanzhou Center for Theoretical Physics, Key Laboratory of Theoretical Physics of Gansu Province, Lanzhou University, Lanzhou, Gansu 730000, China}

\author{Lei Tan}
\email{tanlei@lzu.edu.cn}
\affiliation{Lanzhou Center for Theoretical Physics, Key Laboratory of Theoretical Physics of Gansu Province, Lanzhou University, Lanzhou, Gansu 730000, China}
\affiliation{Key Laboratory for Magnetism and Magnetic Materials of the Ministry of Education, Lanzhou University, Lanzhou 730000, People's Republic of China}


\begin{abstract}
We propose an experimentally feasible scheme for realizing quantum state transfer via the topological edge states in a one-dimensional cavity-magnon lattice. We find that the cavity-magnon system can be mapped analytically into the generalized Su-Schrieffer-Heeger model with tunable cavity-magnon coupling. It can be shown that the edge state can be served as a quantum channel to realize the photonic and magnonic state transfers by adjusting the cavity-cavity coupling strength. Further, our scheme can realize the quantum state transfer between photonic state and magnonic state by changing the amplitude of the intracell hopping. With a numerical simulation, we quantitatively show that the photonic, magnonic and magnon-to-photon state transfers can be achieved with high fidelity in the cavity-magnon lattice. Spectacularly, the three different types of quantum state transfer schemes can be even transformed to each other in a controllable fashion. This system provides a novel way of realizing quantum state transfer and can be implemented in quantum computing platforms.
\end{abstract}

\maketitle
\section{INTRODUCTION}

Quantum state transfer (QST) is of major interest in physics with potential applications in large-scale quantum information processing [\onlinecite{Saffman1, Duan2,Suter3,Kay03,Paganelli031,Lorenzo032,Plenio033,Silveri034,Vijay035, Stannigel035,Chen036,Zhang037,Li038,Stannigel039,Rips040,Chen041,Cirac4,Christandl5,Wang6,Zhang7,Zheng8,
Yang9,Yang10,Bose11,Yao12,Brandes13,He14}]. Several schemes have been proposed to discuss QST theoretically and experimentally. Such as, cavity electromechanical system [\onlinecite{Zheng8,Yang9,Yang10}], the spin chain [\onlinecite{Paganelli031},\onlinecite{Lorenzo032},\onlinecite{Christandl5},\onlinecite{Bose11},\onlinecite{Yao12}], quantum dots [\onlinecite{Brandes13},\onlinecite{He14}] and the trapped ions system [\onlinecite{Duan2}], etc. Among these studies, it can be found that the major requirement of QST protocols is robustness. To this regard, the topological QST has attracted much attention because the topology can provide invariance against sizable imperfections in quantum information processing [\onlinecite{Hasan26, Qi27,Chiu28,Bansil29,Shen29,Wu30,Chen31,Paananen32,Malki33,Brouwer34}]. The Su-Schrieffer-Heeger (SSH) model is a promising platform to analyze the topological QST due to its structural simplicity and abundant physical insight [\onlinecite{Li16,Liberto18,Bao19,Guo20}]. Concretely, the robust QST assisted by Landau-Zener tunneling has been reported [\onlinecite{Longhi66}]. In addition, Ref. [\onlinecite{Dlaska2017}] has displayed the robust QST via topologically protected edge pumping in two-dimensional topological spin system. Subsequently, in superconducting qubit chain [\onlinecite{Mei07}] and superconducting flux qubit chain [\onlinecite{Zheng072}], robust QST via topological edge states has been realized. Especially, Qi $et$ $al.$ [\onlinecite{Qi20}] studied the one-dimensional modulated SSH chain composed of small optomechanical lattice to implement photonic and phononic topological state transfer.

On the other hand, cavity optomagnonic systems have been extensively investigated recently, which provide a new platform for studying quantum effect [\onlinecite{Dany35,Zhao36,Osada37,Gao38,Haigh39,Zhang40,Zhang41,Tabuchi42,Goryachev43,Li44}]. Magnon, the collective spin excitation for ferromagnetic material, can realize the strong (even ultrastrong) couplings with cavity photons [\onlinecite{Dany35}]. Further, the yttrium iron garnet (YIG) sphere is often used as the ferromagnetic material for the experiment, and this has been characterized by high collective spin excitation density and low dissipation [\onlinecite{Chumak45},\onlinecite{Tan46}]. In addition, the strong coupling between cavity photons and magnons has been observed at both low, room and high temperature experimentally [\onlinecite{Xiao47},\onlinecite{Liu89}]. Moreover, compared to cavity-optomechanical systems, the cavity optomagnonic system possesses a long coherence time and intrinsically good tunability [\onlinecite{Dany35}]. Based on the merits of the cavity optomagnonic system, it is also expected to implement the potential application in
quantum information networks [\onlinecite{Li44}], quantum sensing [\onlinecite{Cao47},\onlinecite{Ebrahimi48}] and magnon dark modes in magnon-gradient memory [\onlinecite{Zhang49}].

Here, in terms of these excellent properties of cavity magnetic systems, a new protocol for implementing QST can be proposed. By putting YIG spheres into the cavity, a one-dimensional cavity magnetic chain is constructed. We analytically find that the system is equivalent to the generalized SSH model. Then, through adiabatical ramping of the cavity-magnon couplings, we find that the topological edge states can be obtained and used as topologically protected quantum channels to realize the photonic, magnonic and magnon-to-photon state transfers. Our results numerically demonstrate that the process of QST is accomplished with high fidelity via the gap state because of the topological protection. More spectacularly, we unveil that the different kinds of transition channels can be switched to each other. Our results provide an experimentally feasible scheme for implementing the QST in a cavity-magnon system, which will offer potential application for quantum information processing.

The paper is organized as follows. Sec. \ref{II} presents the theoretical model of cavity-magnon chain which can be mapped into a tight-binding SSH model. Sec. \ref{III} is devoted to research the photonic state transfer in the cavity-magnon chain. Sec. \ref{IV} explores the magnonic state transfer and magnon-to-photon state transfer. Finally, the conclusions are presented in Sec. \ref{V}.

\section{THE cavity-magnon chain}\label{II}
\begin{figure}[!htbp]
\includegraphics[width=7.2cm,height=3.2cm]{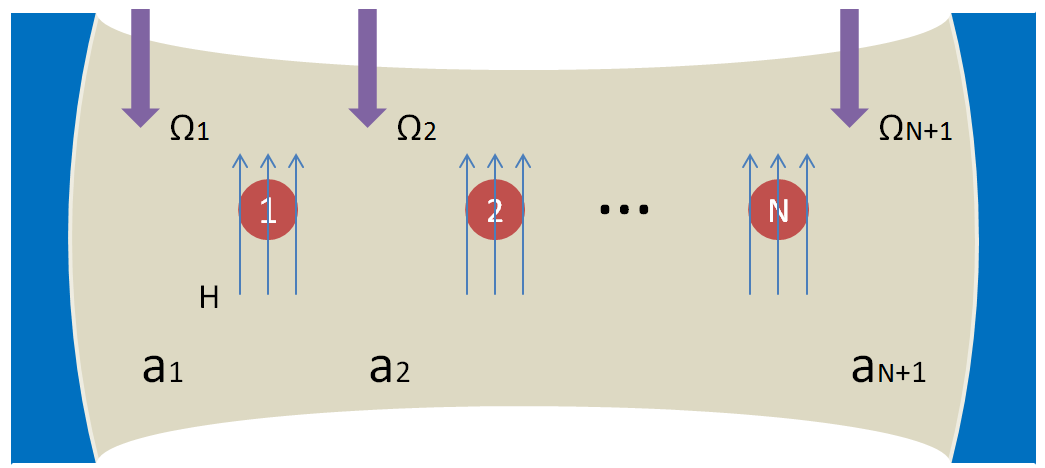}
\caption{(Color online) The diagrammatic sketch of a one-dimension cavity-magnon system consisting of $N+1$ cavity modes and $N$ YIG spheres. $a_{1}$, $a_{2}$... and $a_{N+1}$ represent cavity modes which is pumped by the laser shown as purple arrows. YIG spheres (red spheres) are placed at the cavity and in a uniform bias magnetic field. The coupling between magnon $m_{n}$ and the two adjacent cavity is $g_{n}$ and $g_{n}^{'}$, respectively. In addition, the two adjacent cavities have the coupling with the coupling strength $J$.}
\label{fig1}
\end{figure}
We consider a one-dimensional cavity-magnon system composed $N+1$ cavity modes [\onlinecite{Strackcavity,Sundaresancavity,Torgglercavity}] and $N$ YIG spheres (red spheres), as depicted in Fig. \ref{fig1}. Each cavity field is driven by a laser with strength $\Omega_{n}$$(n=1,2,...,N+1)$. Specifically, the system
can be described by

\begin{flalign}
H_{1}=&\sum\limits_{n=1}^{N+1}(\omega_{a,n}{a_{n}^\dag}{a_{n}}+\Omega_{n}{a_{n}^\dag}e^{-i\omega_{d}t}+\Omega_{n}^{*}a_{n}e^{i\omega_{d}t})\nonumber\\
&+\sum\limits_{n=1}^{N}\omega_{m,n}{m_{n}^\dag}{m_{n}}\nonumber\\
&+\sum\limits_{n=1}^{N}[g_{n}({a_{n}^\dag}m_{n}+a_{n}{m_{n}^\dag})
+g_{n}^{'}({a_{n+1}^\dag}m_{n}+a_{n+1}{m_{n}^\dag})\nonumber\\
&+J({a_{n+1}^\dag}a_{n}+{a_{n}^\dag}a_{n+1})],
\end{flalign}
where $a_{n}^\dag$ ($a_{n}$) and $m_{n}^\dag$ ($m_{n}$) are the photonic and magnonic creation (annihilation) operators. The first term is the free energy of the cavity fields with frequencies $\omega_{a,n}$ and driving laser with frequency $\omega_{d}$. The second term is the free energy of the magnons with frequencies $\omega_{m,n}$.
The last term represents the coupling between the cavity fields and the magnons through magnetic dipole coupling and cavity-cavity coupling, in which the coupling strengths are $g_{n}$,  $g_{n}^{'}$ and $J$, respectively. Then, using a rotating transformation with the external driving frequency $\omega_{d}$, the total Hamiltonian $H_{1}$ becomes

\begin{flalign}
H_{2}=&\sum\limits_{n=1}^{N+1}(\Delta_{a,n}{a_{n}^\dag}{a_{n}}+\Omega_{n}{a_{n}^\dag}+\Omega_{n}^{*}a_{n})\nonumber\\
&+\sum\limits_{n=1}^{N}\Delta_{m,n}{m_{n}^\dag}{m_{n}}\nonumber\\
&+\sum\limits_{n=1}^{N}[g_{n}({a_{n}^\dag}m_{n}+a_{n}{m_{n}^\dag})
+g_{n}^{'}({a_{n+1}^\dag}m_{n}+a_{n+1}{m_{n}^\dag})\nonumber\\
&+J({a_{n+1}^\dag}a_{n}+{a_{n}^\dag}a_{n+1})],
\end{flalign}
where $\Delta_{a,n}=\omega_{a,n}-\omega_{d}$ ($\Delta_{m,n}=\omega_{m,n}-\omega_{d}$) is the detuning between the cavity fields (magnon modes) and the external lasers. Subsequently, we utilize the mean field approximation method to analyse
the steady-state dynamics of the cavity-magnon lattice. In other words, the operators $a_{n}$ and $m_{n}$ are replaced by  $a_{n}=\langle a_{n}\rangle+\delta a_{n}=\alpha_{n}+\delta a_{n}$ and $m_{n}=\langle m_{n}\rangle+\delta m_{n}=\beta_{n}+\delta m_{n}$, respectively [\onlinecite{Xu100}]. After dropping the notation "$\delta$" for all the fluctuation operators $\delta a_{n}$ ($\delta m_{n}$) [\onlinecite{Qi20}], the
Hamiltonian is given by
\begin{flalign}
H_{3}=&\sum\limits_{n=1}^{N+1}\Delta_{a,n}{a_{n}^\dag}{a_{n}}\nonumber\\
&+\sum\limits_{n=1}^{N}\Delta_{m,n}{m_{n}^\dag}{m_{n}}\nonumber\\
&+\sum\limits_{n=1}^{N}g_{n}[({a_{n}^\dag}m_{n}+a_{n}{m_{n}^\dag})
+g_{n}^{'}({a_{n+1}^\dag}m_{n}+a_{n+1}{m_{n}^\dag})\nonumber\\
&+J({a_{n+1}^\dag}a_{n}+{a_{n}^\dag}a_{n+1})].
\end{flalign}
We further implement another rotating transformation with respect to $\Delta_{a,n}{a_{n}^\dag}{a_{n}}$ and $\Delta_{m,n}{m_{n}^\dag}{m_{n}}$, i.e.,
\begin{equation}
U=\exp{[-i(\sum\limits_{n=1}^{N+1}\Delta_{a,n}a_{n}^\dag}{a_{n}+\sum\limits_{n=1}^{N}\Delta_{m,n}m_{n}^\dag}{m_{n})]}.
\end{equation}
Then, the Hamiltonian becomes
\begin{flalign}
H_{4}=&\sum\limits_{n=1}^{N}[g_{n}({a_{n}^\dag}m_{n}e^{i(\Delta_{a,n}-\Delta_{m,n})t}+a_{n}{m_{n}^\dag}e^{-i(\Delta_{a,n}-\Delta_{m,n})t})\nonumber\\
&+g_{n}^{'}({a_{n+1}^\dag}m_{n}e^{i(\Delta_{a,n}-\Delta_{m,n})t}+a_{n+1}{m_{n}^\dag}e^{-i(\Delta_{a,n}-\Delta_{m,n})t})\nonumber\\
&+J({a_{n+1}^\dag}a_{n}+{a_{n}^\dag}a_{n+1})].
\end{flalign}
When the parameters satisfy $\Delta_{a,n}=\Delta_{m,n}$, the Hamiltonian ultimately becomes
\begin{flalign}
H_{5}=&\sum\limits_{n=1}^{N}[g_{n}({a_{n}^\dag}m_{n}+a_{n}{m_{n}^\dag})
+g_{n}^{'}({a_{n+1}^\dag}m_{n}+a_{n+1}{m_{n}^\dag})\nonumber\\
&+J({a_{n+1}^\dag}a_{n}+{a_{n}^\dag}a_{n+1})].
\end{flalign}

Remarkably, if $J=0$, the Hamiltonian $H_{5}$ only possesses the interactions between the adjacent cavity field and the magnonic mode. It means that
the one-dimensional cavity-magnon lattice can be transformed into a generalized SSH model physically.

\section{photonic STATE TRANSFER in Cavity- Magnon chain}\label{III}
\begin{figure}[!htbp]
\centering
\includegraphics[width=4.2cm,height=3.8cm]{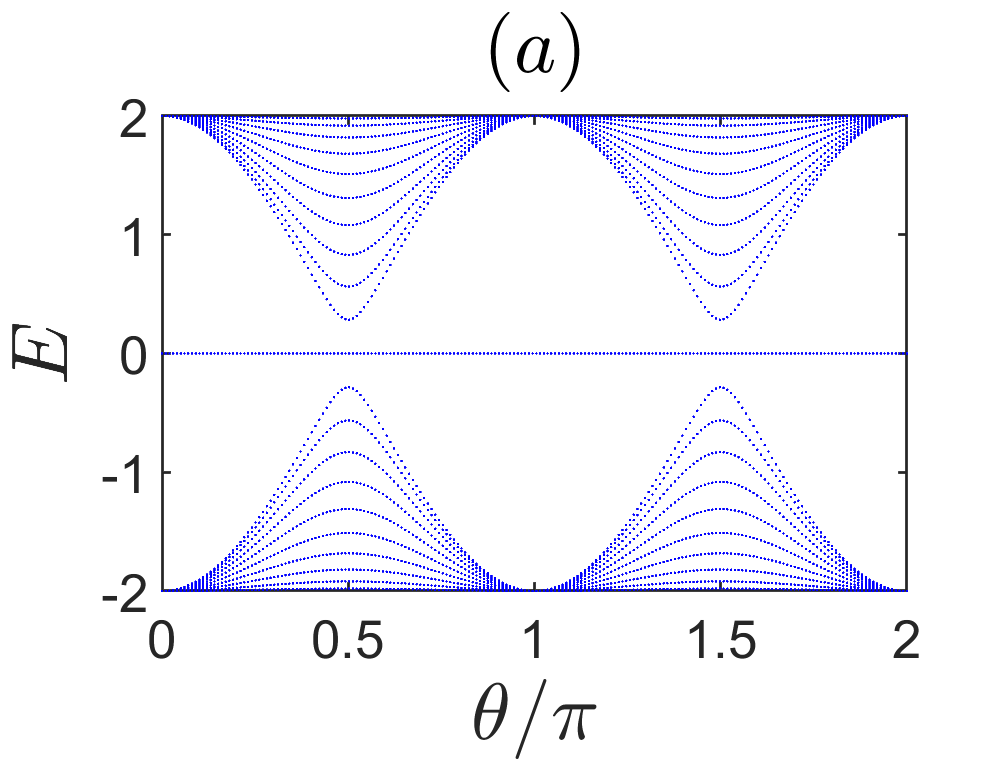}
\includegraphics[width=4.2cm,height=3.8cm]{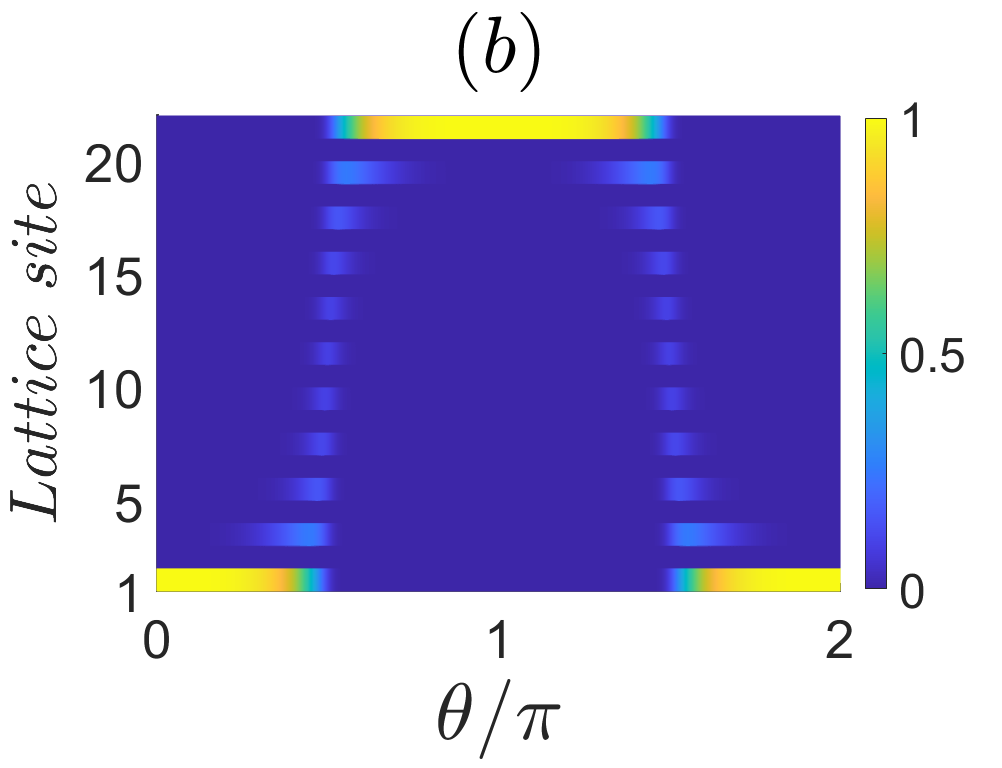}

\includegraphics[width=4.2cm,height=3.8cm]{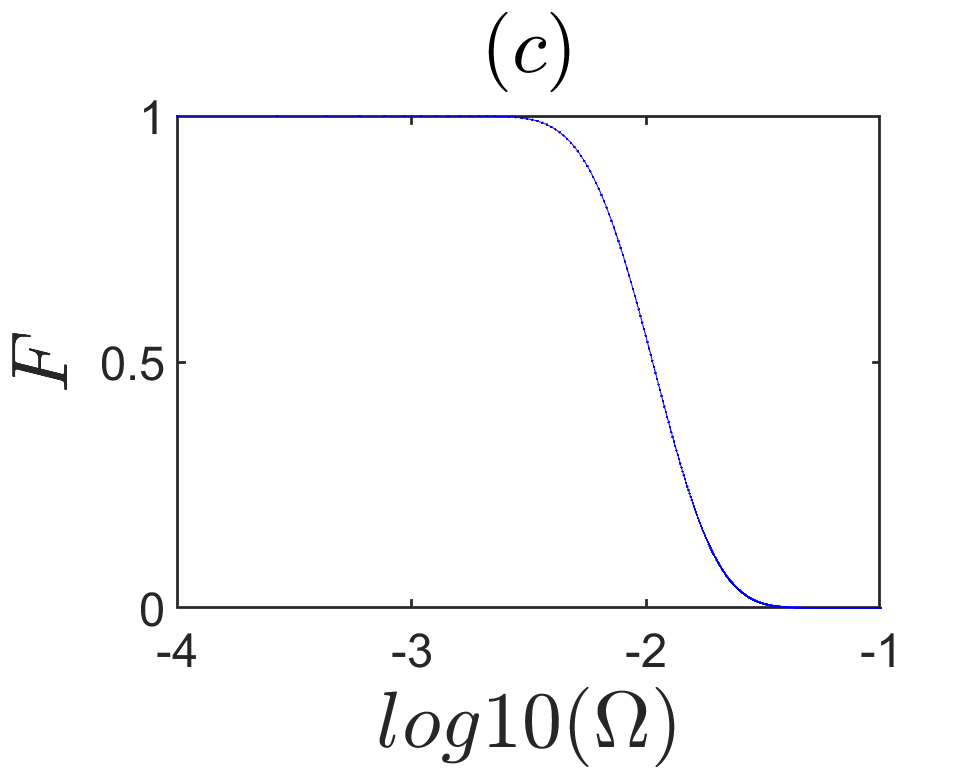}
\caption{(Color online) (a) The energy spectra of the cavity-magnon system with $g_{0}=g_{0}^{'}=1$ and $J=0$. A zero-energy state always exists. (b) The corresponding state distribution of zero-energy state. (c) The fidelity of the state transfer between $|L\rangle$ and $|R\rangle$ versus the varying rate of $\Omega$.}
\label{fig2}
\end{figure}

We first consider the standard SSH model, i.e., $J=0$. The Hamiltonian becomes
\begin{equation}
H_{6}=\sum\limits_{n=1}^{N}[g_{n}({a_{n}^\dag}m_{n}+a_{n}{m_{n}^\dag})+g_{n}^{'}({a_{n+1}^\dag}m_{n}+a_{n+1}{m_{n}^\dag}),
\end{equation}
where we take $g_{n}=g$ and $g_{n}^{'}=g^{'}$ and both $g$ and $g^{'}$ in a periodic way as $g=g_{0}(1-\cos\theta)$ and $g^{'}=g_{0}^{'}(1+\cos\theta)$. The edge state are exponentially localized at the boundaries. Specifically, in single excitation subspace, the wave function has the form $|\Psi \rangle(\theta)=\sum_{n}\lambda^{n}(\alpha a_{n}^\dag+\beta m_{n}^\dag)|G \rangle$ [\onlinecite{Mei07}], in which $|G \rangle=|0,0,0,...0 \rangle$ and $\lambda$ denotes the localization indexes. Then, the eigenvalue equation can be acquired as
\begin{flalign}
g \lambda^{n} \beta a_{n}^\dag|G \rangle &+ g^{'} \lambda^{n+1} \alpha m_{n}^\dag |G \rangle +g \lambda^{n} \alpha m_{n}^\dag |G \rangle + g^{'} \lambda^{n-1} \beta a_{n}^\dag |G \rangle\nonumber\\
&=E\lambda^{n}(\alpha a_{n}^\dag+\beta m_{n}^\dag).
\end{flalign}
Therefore, the zero-energy edge state wavefunction can be derived as
\begin{equation}
|\Psi\rangle=\sum_{n} (-\frac{g}{g^{'}})^{n} a_{n}^\dag |G\rangle,
\end{equation}
Eq. (9) analytically demonstrates that the edge state occupies near the left boundary when $\theta \in (0,\frac{\pi}{2})$ and $\theta \in (\frac{3\pi}{2},2\pi)$, while $\theta \in (\frac{\pi}{2},\frac{3\pi}{2})$, it is localized near the right boundary. Note that both the left and right sites belong to the cavity fields. To further visualize the above analytical expression, we plot the distribution of the zero-energy mode, as shown in Fig. \ref{fig2}(b). One can find that the numerical results are consistent well with the analytical calculation. Then, the result above provides some important information that we can achieve a photonic topological state transfer between the first and the last cavity fields assisted by the zero-energy mode via varying the periodic parameter $\theta$ from $0$ to $\pi$. Concretely, the cavity-magnon coupling strength becomes $g=0$ ($ g^{'}=0$) when $\theta=0$ ($\theta=\pi$), which induces that the leftmost (rightmost) cavity mode is decoupled from the rest of the cavity-magnon chain. Therefore, the edge states become

\begin{equation}
|L \rangle=|1,0,0...0 \rangle,
\end{equation}
\begin{equation}
|R \rangle=|0,0,0...1 \rangle.
\end{equation}

Here, the photon state can be transferred adiabatically via the channel of the zero-energy edge state. To do this, the periodic parameter $\theta$ is set as $\theta(t)=\Omega t$, with $\Omega $ being the ramping frequency [\onlinecite{Mei07}]. If the initial state is prepared in the photonic left edge state $|L\rangle=|1,0,0,...,0\rangle$, the photonic right edge state can be obtained through the evolution of the time-dependent Hamiltonian $i\frac{\partial}{\partial_{t}}|\Psi_{t} \rangle=H(\theta_{t})|\Psi_{t}\rangle$. Further, the fidelity of this evolution process $F=|\langle R|\Psi_{f}\rangle|$ can be numerically calculated in Fig. \ref{fig2}(c). Affirmatively, the fidelity is maintained as the value of unity over a large range. For example, we can choose $\Omega=0.03g_{0}^{'}$ to ensure the photonic state transfer from $|L \rangle$ to $|R \rangle$ with high fidelity. The coupling strength of cavity-magnon can reach $g/2\pi=g^{'}/\pi=2$ GHz [\onlinecite{Goryachev43}]. Namely, this strong coupling strength corresponds to the time of QST being $t=\pi/\Omega=8.3$ ns, which is fast.

\begin{figure}[!htbp]
\centering
\includegraphics[width=4.2cm,height=3.8cm]{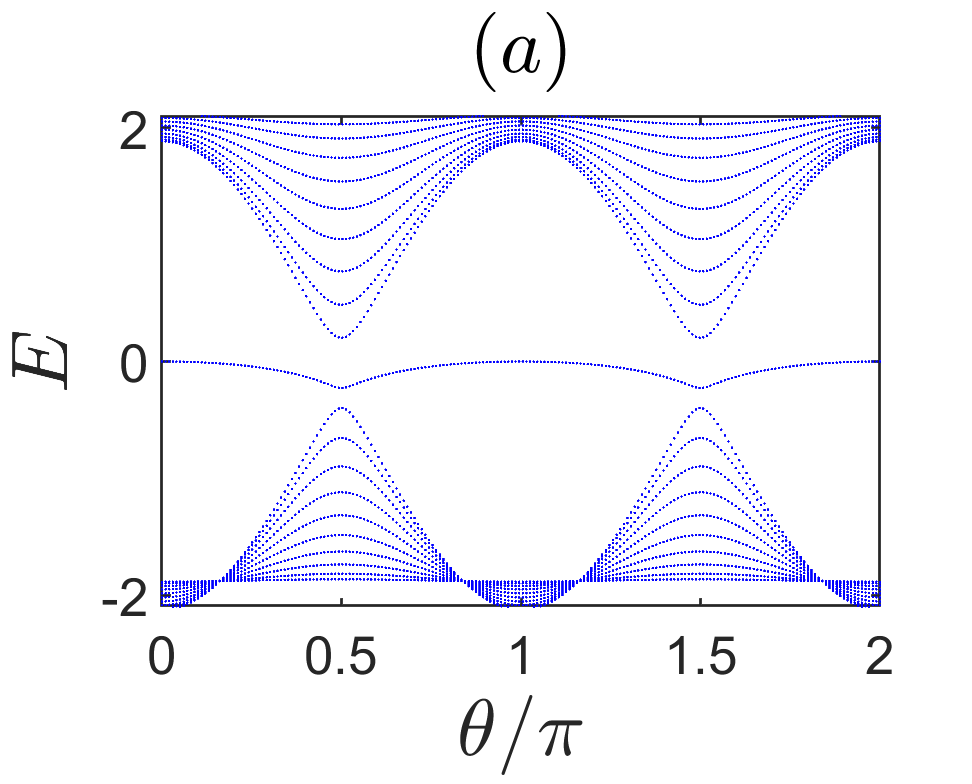}
\includegraphics[width=4.2cm,height=3.8cm]{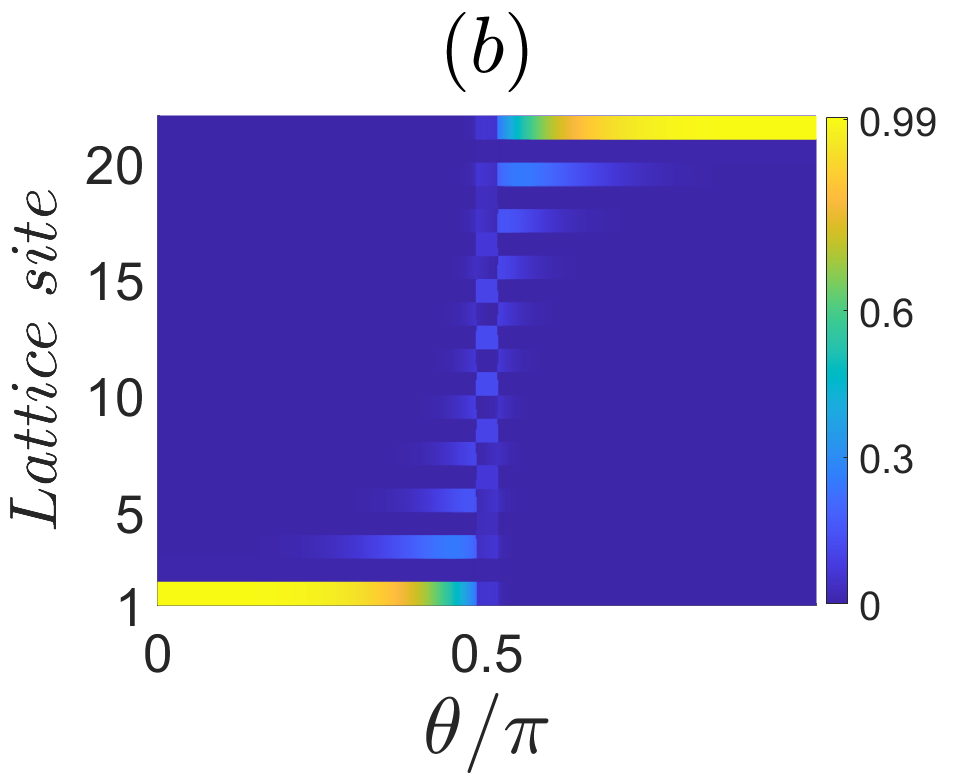}

\includegraphics[width=4.2cm,height=3.8cm]{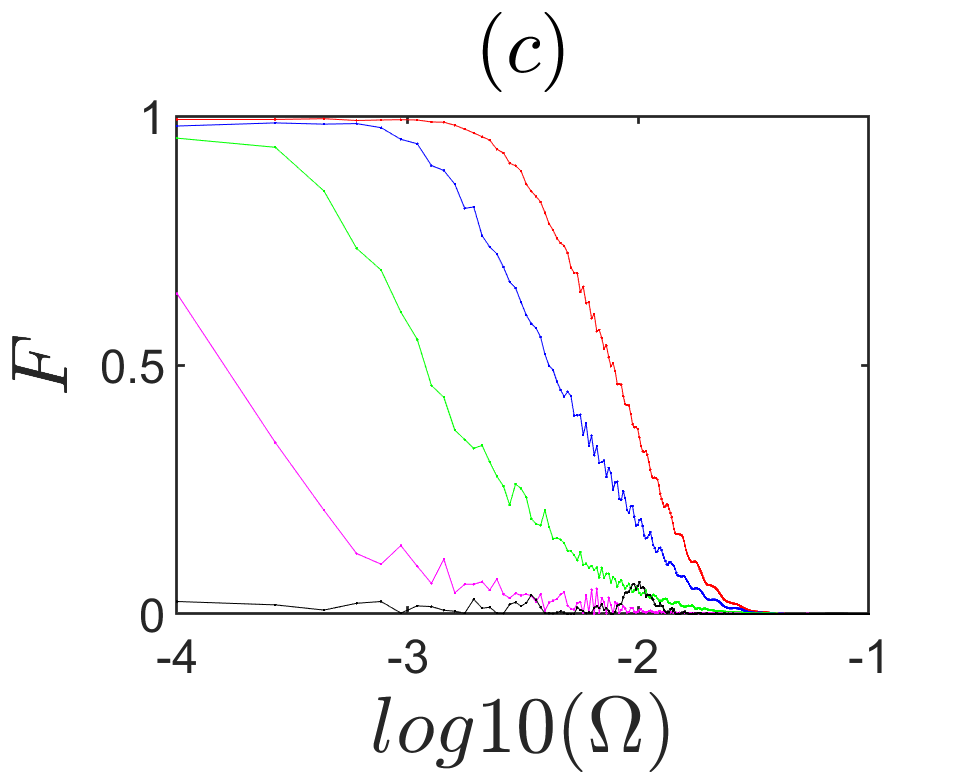}
\includegraphics[width=4.2cm,height=3.8cm]{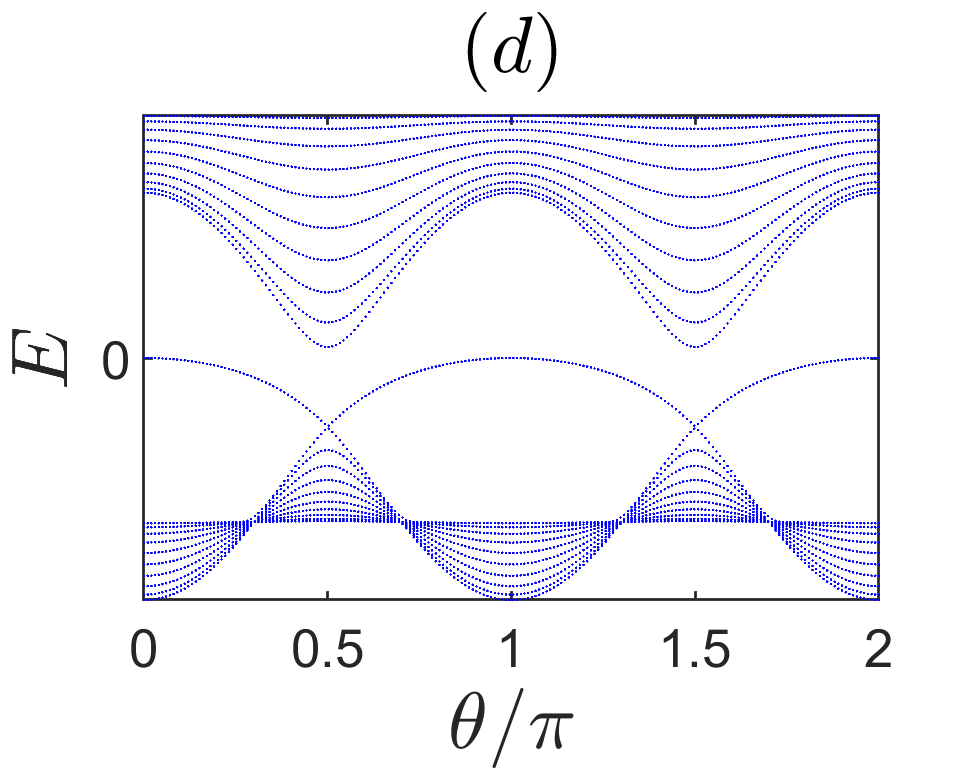}

\caption{(Color online) (a) The energy spectra of the cavity-magnon system with $g_{0}=g_{0}^{'}=1$ and $J=0.125$. A gap state always exists. (b) The corresponding state distribution. (c) The fidelity with $J=0.125$(the red line), $J=0.2$(the blue line), $J=0.3$(the green line), $J=0.4$(the pink line) and $J=0.5$(the black line). (d) The energy spectrum with $J=0.5$. Here, the lattice size is $L=21$.}
\label{fig3}
\end{figure}

Further, we consider the case of $J\neq0$. In Figs. \ref{fig3}(a) and \ref{fig3}(b), the energy spectrum and the corresponding gap state distribution can be displayed with $J=0.125$. Clearly, this gap state is not the zero mode, but the energy gap is still nonzero. In other words, this state also can be viewed as a channel to realize QST. Correspondingly, as shown in Fig. \ref{fig3}(b), the photon prepared initially in the first cavity mode can be transferred into the last cavity mode finally. In Fig. \ref{fig3}(c), we exhibit the fidelity of this evolution process once again, in which the red line stands for $J=0.125$. Obviously, the fidelity $F>99\%$ when $-4<log_{10}{\Omega}<-2.5$, i.e., the photon state can be transferred almost perfectly. In addition, we also calculate the fidelity for some values of $J$. It can be shown that the situation becomes different when $J$ continues to increase but less than $g$. Concretely, with $J=0.5$ (the black line), the fidelity is close to zero. Physically, it can be explained as that the large $J$ completely destroys the original state transfer channel, as shown in Fig. \ref{fig3}(d), with the gap states coming into the bulk.

Up until we found that the photonic state can be transferred across over the chain with high fidelity. Next, we will explore that what new phenomena will occur in our cavity-magnon system.

\section{MAGNONic STATE TRANSFER and MAGNON-TO-PHOTON STATE TRANSFER }\label{IV}
In this section, we now discuss other state transfer protocols in cavity-magnon system by changing the cavity-cavity coupling strength ($J$) and the intracell hopping strength ($g_{0}$).

\begin{figure}[!htbp]
\centering
\includegraphics[width=4.2cm,height=3.8cm]{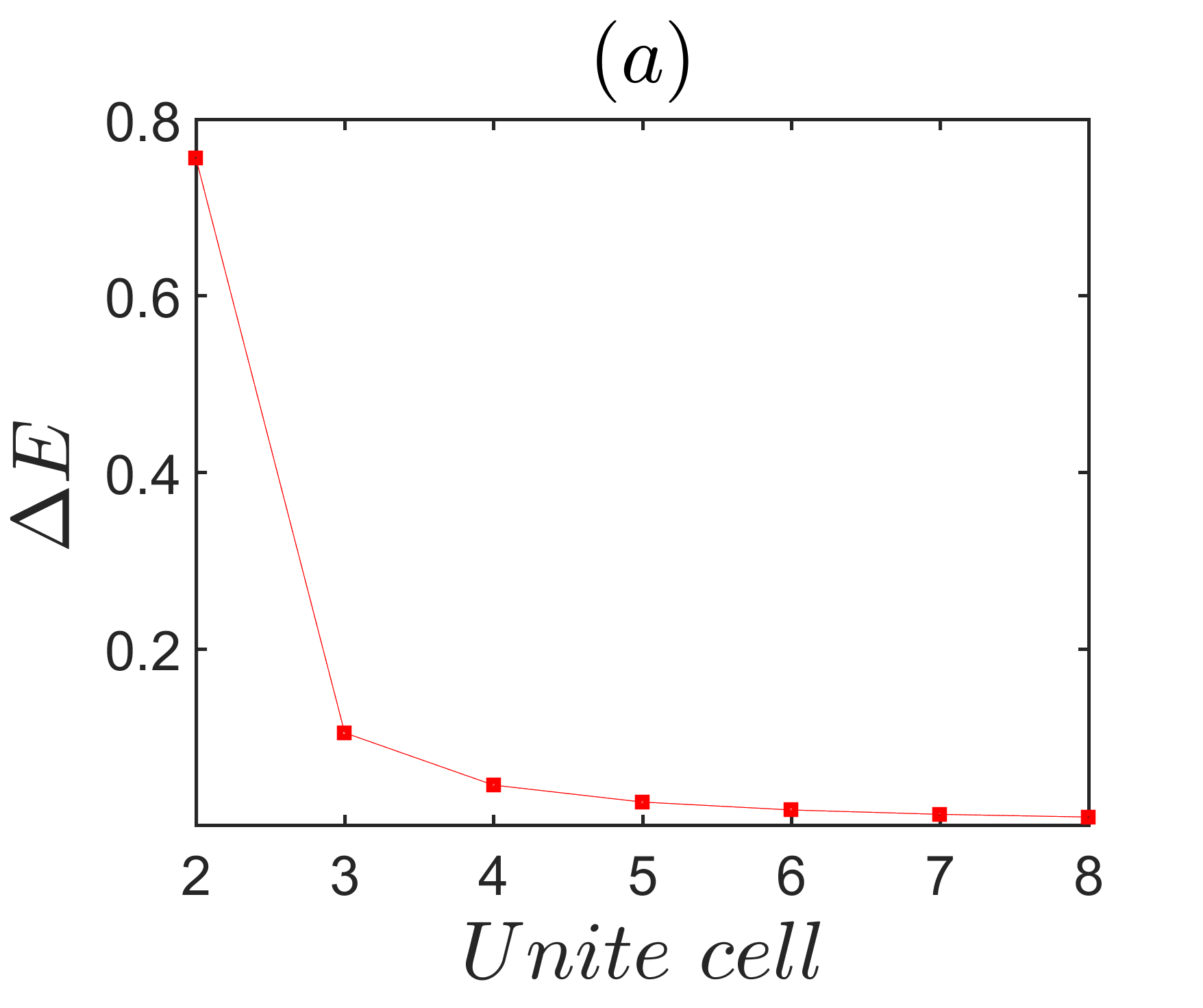}

\includegraphics[width=4.2cm,height=3.8cm]{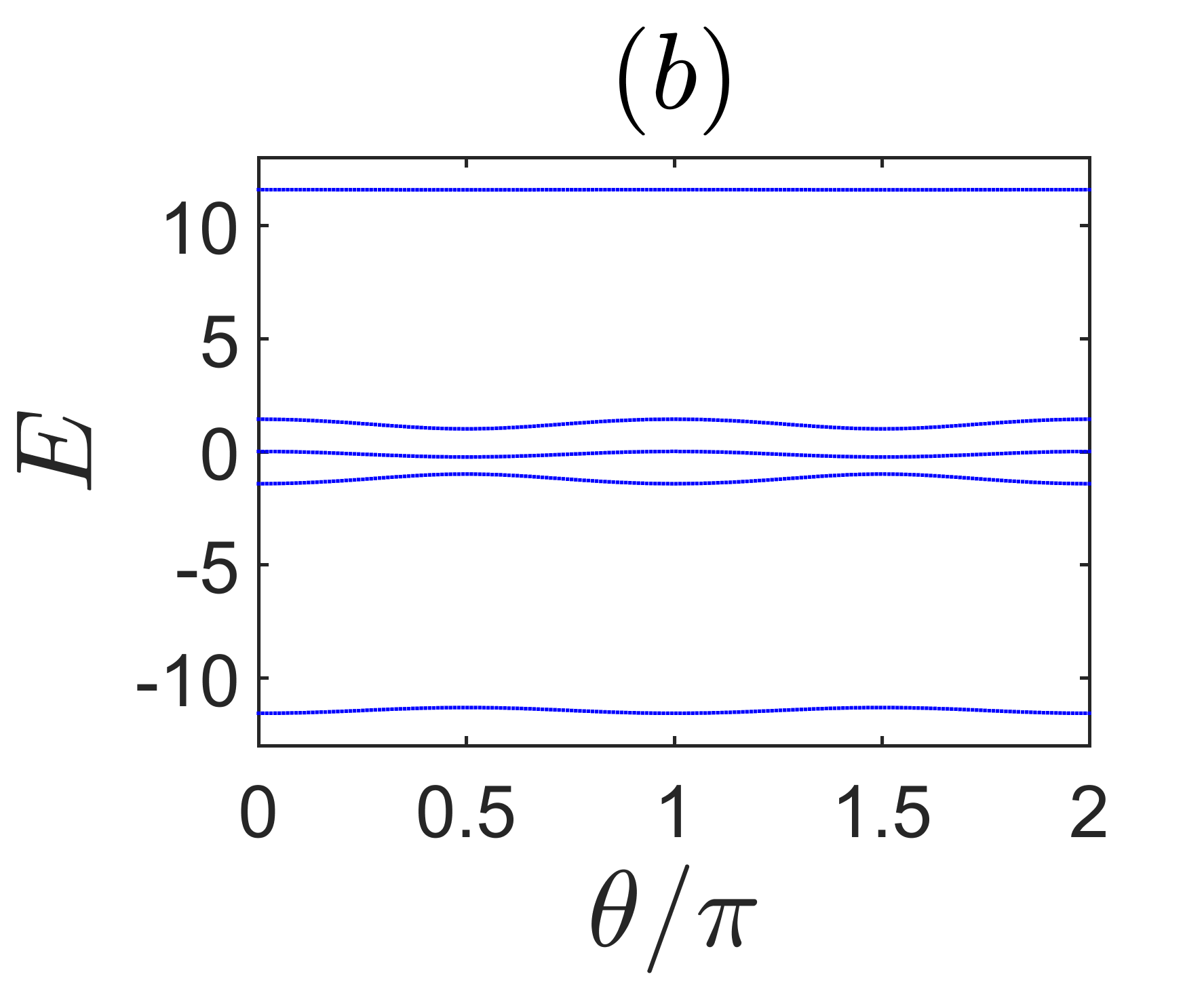}
\includegraphics[width=4.2cm,height=3.8cm]{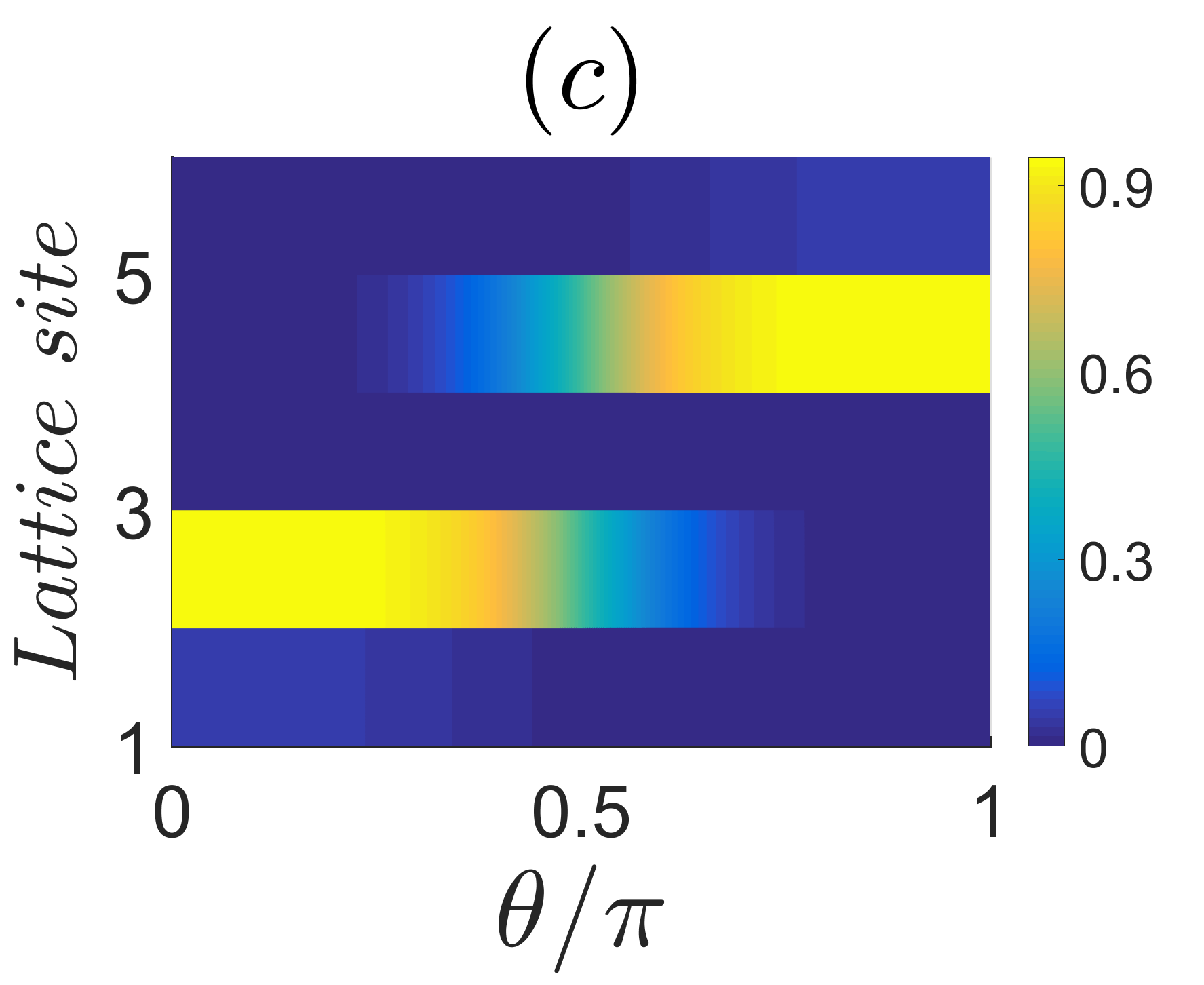}

\includegraphics[width=4.2cm,height=3.8cm]{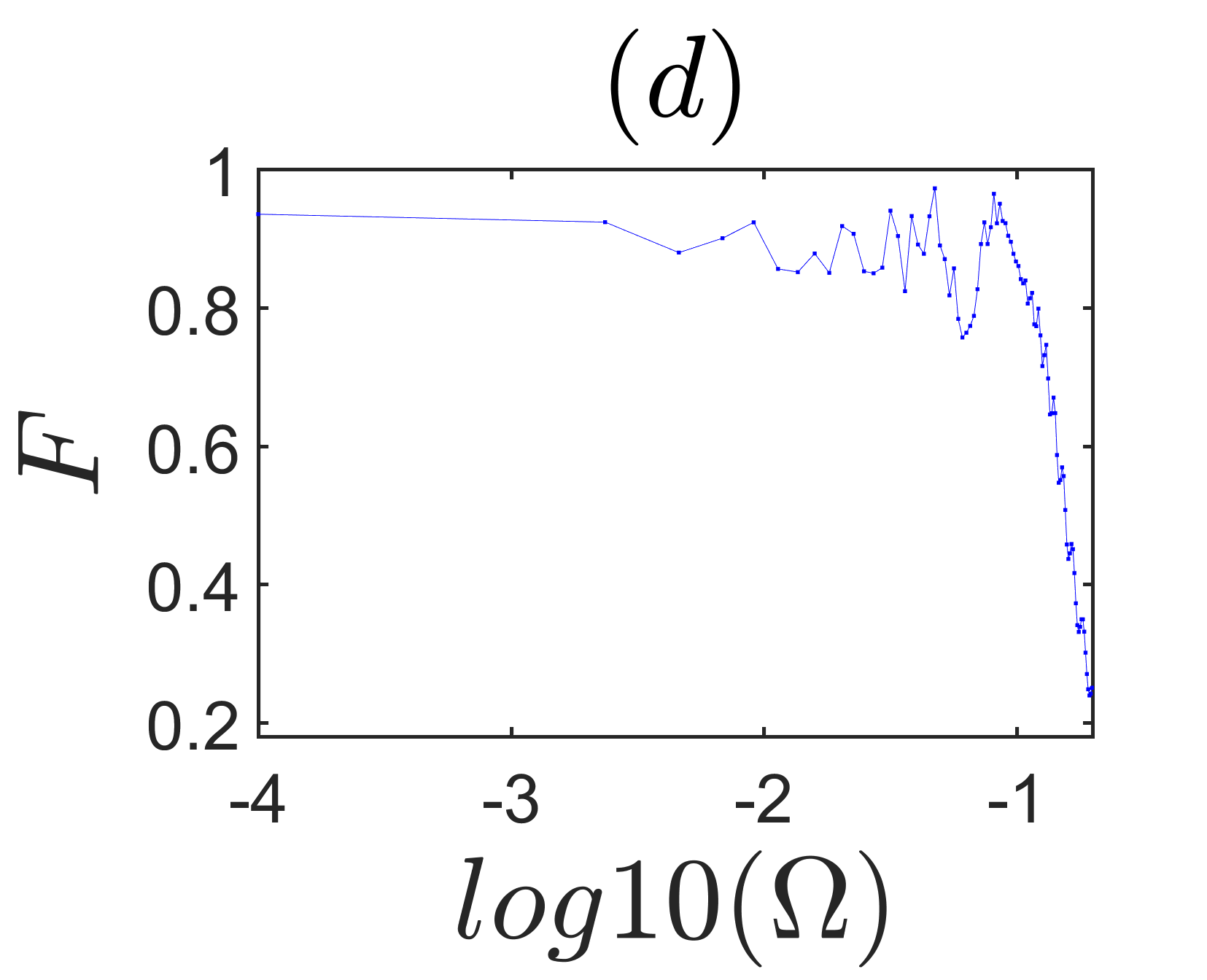}
\includegraphics[width=4.2cm,height=3.8cm]{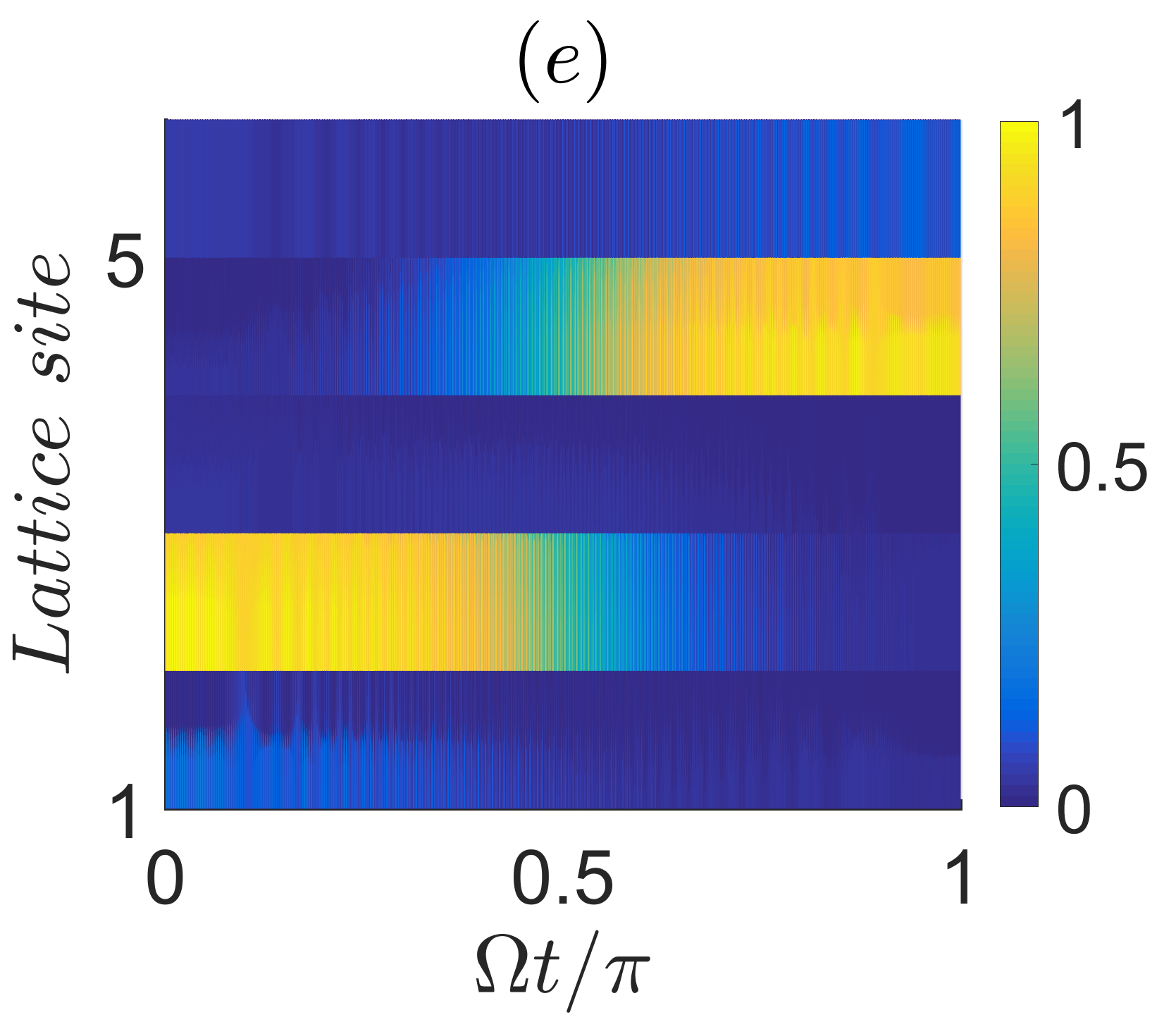}

\caption{(Color online) (a) The energy gap (the gap state and the low band) $\Delta_{E}$ versus the length of the chain. (b) The energy spectra of the cavity-magnon system. It always exist a gap state. (c) The corresponding state distribution of gap state. (d) The fidelity between $|\psi_{0} \rangle=|0,1,0,0,0 \rangle$ and $|\psi_{\pi} \rangle=|0,0,0,1,0 \rangle$. (e) The state transfer process corresponding to $\Omega=3\times10^{-4}$. Other parameters are $g_{0}=g_{0}^{'}=1$ and $J=8$.}
\label{fig4}
\end{figure}

$Magnonic$ $state$ $transfer$ As we all know, the closure of the energy gap marks that the gap state is not existence, i.e., the choice is not suitable for QST. In Fig. \ref{fig4} (a), we plot the energy gap $\Delta_{E}$ versus the length of the chain with $g_{0}=g_{0}^{'}=1$ and $J=8$. One can find that the energy gap tends to be narrow as increasing the number of the unit cell. Therefore, in order to better elucidate the state transition process, the length of the following cavity-magnon lattice can be taken as $L=5$.

To explore the QST, as shown in Fig. \ref{fig4}(b), the energy spectrum is obtained. It can be shown that the shape of the gap state is wavy-like, which implies that a new channel realizing another type of state transfer may be apparent. In order to demonstrate this statement, we plot the state distribution in Fig. \ref{fig4}(c). Obviously, corresponding $\theta=0$, the gap state is localized near the first YIG sphere, while it is located near the second YIG sphere when $\theta=\pi$. Those results show that the gap state becomes a new state transfer channel between the magnonic state of $|\psi_{0} \rangle =|0,1,0,0,0 \rangle$ and $|\psi_{\pi} \rangle=|0,0,0,1,0\rangle$. The numerical results mentioned above can be understood physically as follows. The two adjacent cavities coupling strength $J$ is larger, which makes the original three cavity modes can be regarded as a cavity field [\onlinecite{Qi20}]. Therefore, this system can realize state transfer between magnons. To further confirm it, we numerically simulate the fidelity of the magnetic state transfer in Fig. \ref{fig4}(d). We find that, when the ramping speed $\log_{10}\Omega<-2.3$, the state transfer between $|\psi_{0} \rangle=|0,1,0,0,0 \rangle$ and $|\psi_{\pi} \rangle=|0,0,0,1,0 \rangle$ can be realized with high fidelity. We also depict the state transfer process when $\Omega=3\times10^{-4}$ in Fig. \ref{fig4}(e). Obviously, the state transfer between $|\psi_{0} \rangle=|0,1,0,0,0 \rangle$ and $|\psi_{\pi} \rangle=|0,0,0,1,0 \rangle$ can indeed be implemented.

\begin{figure}[!htbp]
\centering
\includegraphics[width=4.2cm,height=3.8cm]{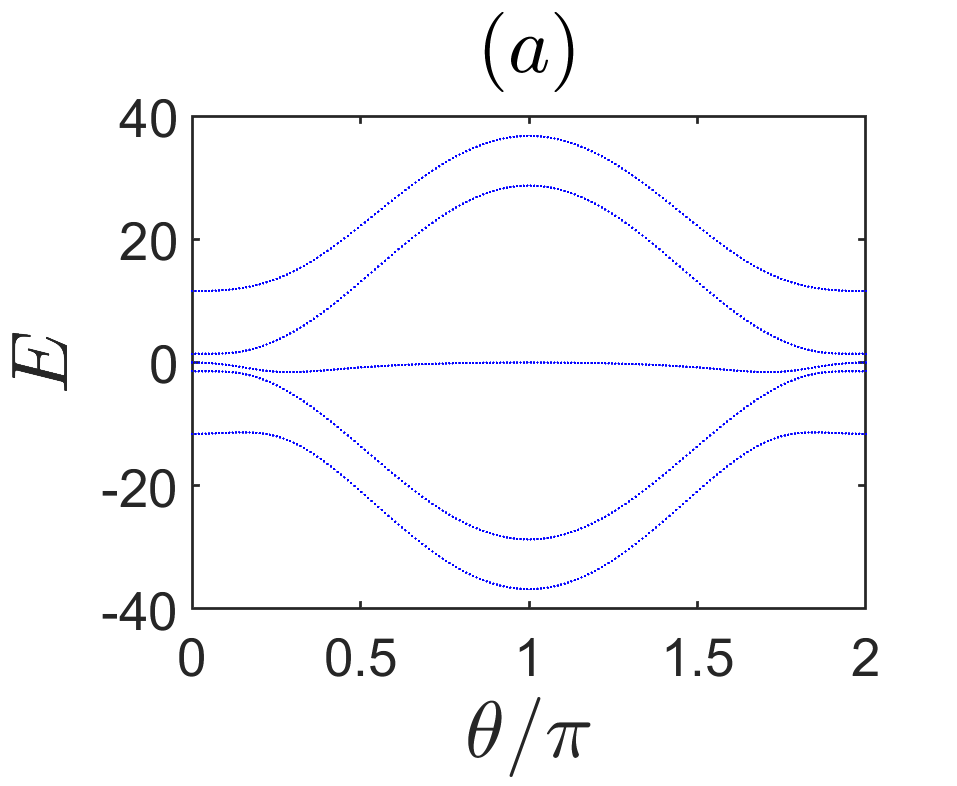}
\includegraphics[width=4.2cm,height=3.8cm]{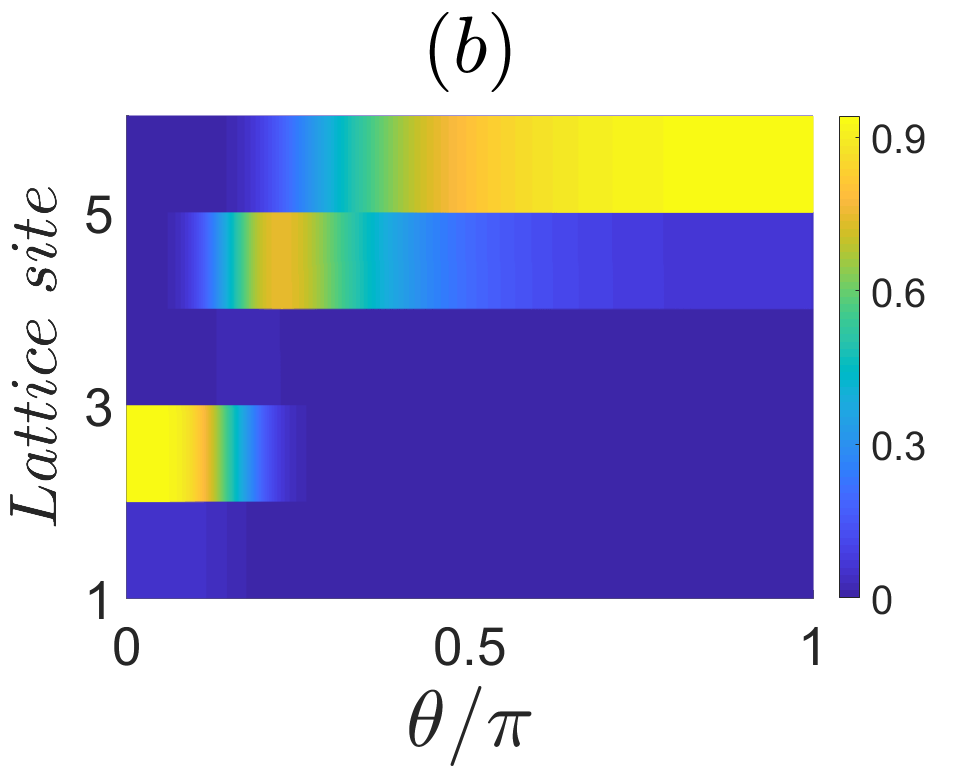}

\includegraphics[width=4.2cm,height=3.8cm]{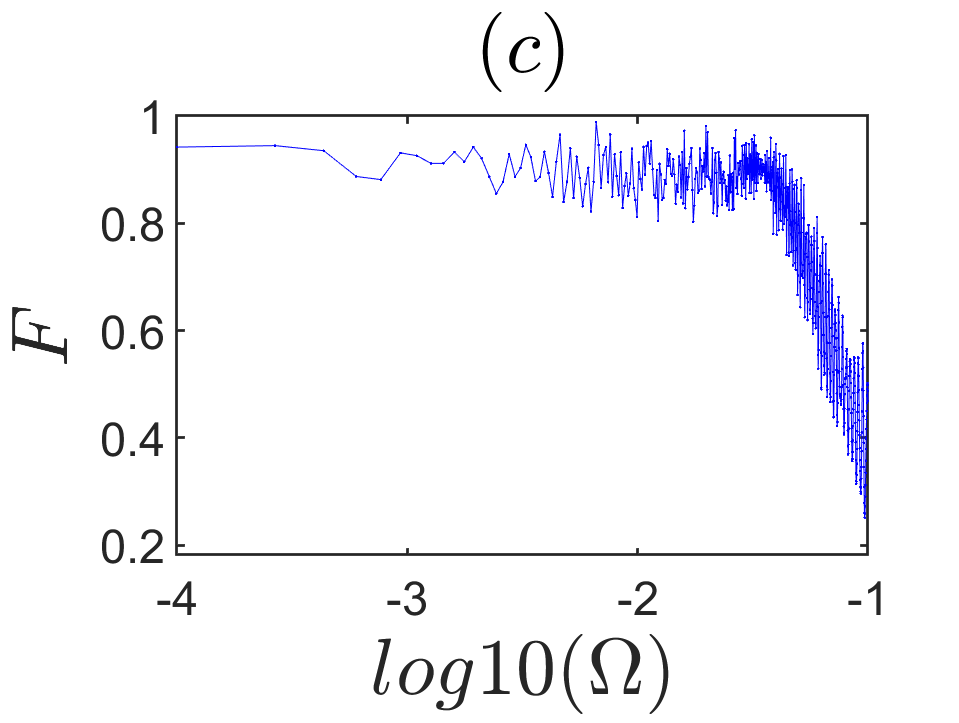}
\includegraphics[width=4.2cm,height=3.8cm]{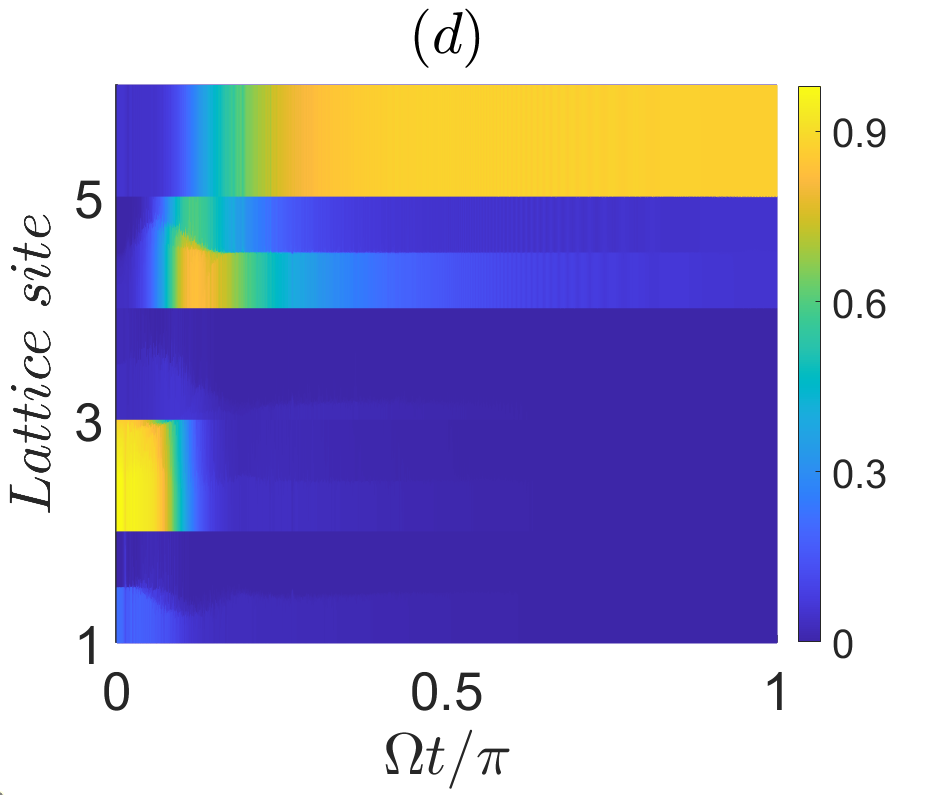}

\caption{(Color online) (a) The energy spectra of the cavity-magnon system with $g_{0}=16$, $g_{0}^{'}=1$ and $J=8$. (b) The corresponding state distribution of gap state. (c) The fidelity between $|\psi_{0} \rangle=|0,1,0,0,0 \rangle$ and $|\psi_{\pi} \rangle=|0,0,0,0,1 \rangle$. (d) The state transfer process corresponding to $\Omega= 0.001$.}
\label{fig5}
\end{figure}

$Magnon-to-photon$ $state$ $transfer$ We have seen that the magnonic state transfer is applicable when $g_{0}=g_{0}^{'}=1$ and $J=8$. Then, what will happen if we change the value of $g_{0}$? We can now take $g_{0}=16$. The energy spectrum and the corresponding gap state distribution are plotted in Figs. \ref{fig5}(a) and \ref{fig5}(b). One can find that the gap state is occupied near the second site when $\theta=0$, whereas it is localized near the last site when $\theta=\pi$. In other word, it can realize the transfer from magnonic state $|\psi_{0} \rangle=|0,1,0,0,0 \rangle$ to photon state $|\psi_{\pi} \rangle=|0,0,0,0,1 \rangle$ by dent of this channel. Similarly, the fidelity of the state transfer versus the ramping speed $\Omega$ can be shown in Fig. \ref{fig5}(c). It can be found that when $-4<\log_{10}\Omega<-3$, the fidelity $F>90\%$. A small $\Omega$ is required to meet the adiabatic evolution condition since the energy gap is narrow at this point. Moreover, to make the results more intuitive, in Fig. \ref{fig5}(d), we choose an appropriate value of $\Omega = 0.001$. The numerical result reveals exactly that the state transfer between $|\psi_{0} \rangle=|0,1,0,0,0 \rangle$ and the photon state $|\psi_{\pi} \rangle=|0,0,0,0,1 \rangle$.

From above all, it seems that although photonic, magnonic and magnon-to-photon state transfers are different state transfer processes. However, all of those processes are fulfilled by the gap state. The information from Sec. \ref{III} and \ref{IV} tell us that the photonic state and magnonic state transfer processes only adjust the cavity-cavity strength $J$ appropriately. Analogously, the magnonic and magnon-to-photon state transfer processes can regulate the intracell hopping $g_{0}$. In other words, the three different types of quantum state transfer schemes can be switched to each other by designing parameters suitably for our system.

\section{CONCLUSION}\label{V}

In this work, based on a new platform of the cavity-magnon system, we explore different types of quantum state transfer in terms of the topologically protected edge state. We analytically find that the cavity-magnon lattice is equivalent to the generalized SSH model. It can be found that this edge state can be employed as a quantum channel to realize the photonic, magnonic state transfers by adjusting the cavity-cavity coupling strength. Further, by changing the intracell hopping amplitude, our scheme can also realize the transfer between photonic state and magnetic state. In addition, one can find that the photonic, magnonic and magnon-to-photon state transfer can be achieved with high fidelity. The results obtained here provide an experimentally feasible scheme for realizing the QST in a cavity magnonic system, which will offer valuable insight for quantum information processing.

Experimentally, it is a mature technology that a strong coupling magnon-photon system can be engineered in experiments [\onlinecite{Tabuchi42},\onlinecite{van80,Wang84,Buks85}]. Moreover, the linear array of $3$D cavities for experiments has been developed, which leads that many YIG spheres couple the cavity fields [\onlinecite{Dubyna87},\onlinecite{FENG88}]. In addition, a multimode cavity couples atom has been realized [\onlinecite{Sundaresancavity}]. Therefor, the theoretical model our proposed may be experimentally realized.

\section{ACKNOWLEDGMENTS}
This work was supported by National Natural Science Foundation of China (Grants No. 11874190, No. 61835013 and No. 12047501), and National Key RD Program of China under grants No. 2016YFA0301500. Support was also provided by Supercomputing Center of Lanzhou University.

\bibliography{refBao}

\end{document}